\documentclass[twocolumn,preprintnumbers,amsmath,aps]{revtex4}
\usepackage{graphicx}
\usepackage{dcolumn}
\usepackage{bm}
\usepackage{color}
\usepackage{amssymb}
\usepackage{array}
\usepackage{multirow}

\begin{document}
\newcommand{\kvec}{\mbox{{\scriptsize {\bf k}}}}
\def\eq#1(\ref{#1)}
\def\fig#1{\ref{#1}}
\preprint{{\it submitted to:} {\bf The European Physical Journal B}}
\title{
---------------------------------------------------------------------------------------------------------------\\Electronic conductance via atomic wires: a phase field matching theory approach}
\author{D. Szcz{\c{e}}{\'s}niak$^{1, 2}$ and A. Khater$^{1}$}
\affiliation{1. Institute for Molecules and Materials UMR 6283, University of Maine, Ave. Olivier Messiaen, 72085 Le Mans, France}
\affiliation{2. Institute of Physics, Jan D{\l}ugosz University in Cz{\c{e}}stochowa, Al. Armii Krajowej 13/15, 42200 Cz{\c{e}}stochowa, Poland}
\email{d.szczesniak@ajd.czest.pl}
%
\date{\today}
\begin{abstract}
A model is presented for the quantum transport of electrons, across finite atomic wire nanojunctions between electric leads, at zero bias limit. In order to derive the appropriate transmission and reflection spectra, familiar in the Landauer-B\"{u}ttiker formalism, we develop the algebraic phase field matching theory (PFMT). In particular, we apply our model calculations to determine the electronic conductance for freely suspended monatomic linear sodium wires (MLNaW) between leads of the same element, and for the diatomic copper-cobalt wires (DLCuCoW) between copper leads on a Cu(111) substrate. Calculations for the MLNaW system confirm the correctness and functionality of our PFMT approach. We present novel transmission spectra for this system, and show that its transport properties exhibit the conductance oscillations for the odd- and even-number wires in agreement with previously reported first-principle results. The numerical calculations for the DLCuCoW wire nanojunctions are motivated by the stability of these systems at low temperatures. Our results for the transmission spectra yield for this system, at its Fermi energy, a monotonic exponential decay of the conductance with increasing wire length of the Cu-Co pairs. This is a cumulative effect which is discussed in detail in the present work, and may prove useful for applications in nanocircuits. Furthermore, our PFMT formalism can be considered as a compact and efficient tool for the study of the electronic quantum transport for a wide range of nanomaterial wire systems. It provides a trade-off in computational efficiency and predictive capability as compared to slower first-principle based methods, and has the potential to treat the conductance properties of more complex molecular nanojunctions.
\end{abstract}
\maketitle
\noindent PACS number(s): 73.63.Nm, 73.63.-b, 03.65.Fd, 31.15.xf

\section{INTRODUCTION}
\label{introduction}

Current technological needs motivate the intensive research of the properties of materials at the nanoscale. One of the most important areas in this respect at present concerns nanoelectronics \cite{agrait}. The great interest in this domain arises from the potential reduction of the size of the circuit components, maintaining their quality and functionality, and aiming at greater efficiency and storage characteristics for physical devices \cite{agrait}, \cite{nitzan}.

Among the various physical components which constitute the nanoelectronic devices, the nanojunctions therein are of crucial importance and will become increasingly so in the future \cite{nitzan}, \cite{lamba}. Such low-dimensional elements, which can act {\it e.g.} as switches or gates \cite{nitzan}, are considered to be key circuit components connecting the larger nanostructures. At present they are experimentally prepared and investigated by means of several techniques, such as the mechanically controllable break junction methods \cite{agrait}, \cite{valkering}, feedback-stabilized break junction technique \cite{smith}, atomic force microscopy \cite{agrait}, scanning tunneling microscopy \cite{agrait}, \cite{kroger}, as well as transmission electron microscopy \cite{jin}, \cite{bettini}.

Atomic wires represent a particular class of such nanojunctions in nanoelectronic circuits. The conductance properties of atomic wires are the most interesting features of these systems, they depend on the materials used to fabricate them and on their structural properties. For example, the conductance of the monatomic wire nanojunctions does not stay constant or decrease when the length of the wire increases, rather it oscillates as a function of the number of the wire atoms. The periodicity of the oscillations may vary depending on the type of the atomic wire. It has been shown that, in the monovalent wire nanojunctions composed of alkali (Na and Cs) or noble metals (Ag, Au and Cu), the conductance oscillations exhibit the two-atom periodicity \cite{lang1}, \cite{smit}, \cite{lee}, \cite{khomyakov}. In contrast the Al wire nanojunctions present the oscillations with a period of four or six atoms, depending on the considered interatomic distances between consecutive sites \cite{thygesen}, \cite{xu}. As well as in Al wires, the conductance oscillations have been observed in other multivalence wire nanojuctions, as platinum and iridium \cite{vega}, carbon \cite{lang2}, \cite{shen}, and silicon systems \cite{mozos}, \cite{zhou}.

Understanding the electronic transport properties of such nanojunctions is hence of particular importance. The electrons which contribute to the transport via nanojunctions present characteristic wavelengths comparable to the size of these components, leading to quantum coherent effects. The properties of the nanoelectronic device and its  functionality may be greatly affected or even built on such nanojunction quantum effects \cite{ke}, and cannot be described in the framework of the classical regime \cite{havu}. In particular, the transport characteristics are usually described employing the scattering theory for electronic excitations in nanostructures \cite{newton}. The relationship between quantum scattering and the electronic conductance is initially provided by the work of Landauer \cite{landauer}, together with its generalization to the case of multiple scattering as developed by B\"{u}ttiker \cite{buttiker}. The determination of the scattering transmission and reflection probabilities for the electrons as quantum particles in nanostructures is consequently a key concern.

The scattering probabilities can be determined thanks to a number of theoretical methods which have a common objective in the framework of the Landauer-B\"{u}ttiker formalism, and can be employed by integrating in their approach,  the density functional theory (DFT) \cite{egami1}, \cite{egami2} or the tight-binding model (TB) \cite{zhang1}, \cite{nozaki}. The most popular methods  are the matrix Green's function method by Ando \cite{ando} or the non-equilibrium Green's function formalism (NEGF), also known as the Keldysh formalism \cite{keldysh}. However, there has been a debate whether the Ando's formalism is complete \cite{krstic}, and as has been pointed out, the combination of the Keldysh formalism and DFT is still very often computationally demanding \cite{havu}, \cite{zhang1}, motivating intensive research for solutions to this difficulty. An interesting discussion of different types of the optimization techniques may be found in \cite{egami1}, \cite{egami2}, \cite{deretzis}.

In this paper we present a compact and efficient theoretical method to calculate the electronic quantum conductance, in the framework of the Landauer-B\"{u}ttiker formalism, for the class of nanojunctions made up of finite atomic wires of conducting atoms. Our model calculations are based on the method of the phase field matching theory (PFMT), which was originally and intensively developed for the scattering and transport of phonon and magnon excitations in nanostructures \cite{khater1}. While at first sight, the formulation of the PFMT may be expected to be formally similar for different types of excitations, {\it i.e.} replacing here the dynamical equation for phonons or magnons by the Schr\"{o}dinger equation, there are however significant technical dissimilarities which play an important role. For example, we can refer to the fact that in the electron scattering problem one has to consider allowed symmetries of the atomic orbitals, which is not the case for phonon and magnon excitations.

To fully benefit from the PFMT method, our model dynamics are described in the framework of the linear combination of atomic orbitals in the Slater-Koster tight-binding approximation \cite{slater}. The PFMT method, presented in this work, can be classified as a finite difference approximation technique. In comparison, an early discussion of this method can be found in \cite{szczesniak}, \cite{khater2}, using a basic analytical formulation for the elementary problem of the electronic transport across a single chemical defect in a model wire nanojunction.

The combination of PFMT and tight-binding model leads to transparent matrix structures which are numerically solvable in a direct manner, in contrast with the iterative matrix algebra characteristic of the NEGF calculations which require special numerical optimization methods \cite{deretzis}. Furthermore, integrated tight-binding approximation allows us to discretize the entire model calculation in the real-space, and equally important, to make time economies in the numerical computations. Here we note that the tight-binding theory has been used successfully in previous work as a basis for electronic transport calculations, for various low-dimensional systems, such as polymer nanojunctions \cite{rabani}, carbon nanotubes \cite{chen}, and the presently popular graphene systems \cite{wu}. This reinforces our choice of the electronic dynamics model.

As an implementation of the PFMT method we consider in the present work two types of atomic wire nanojunction systems at zero bias limit. In the first step our model calculations are applied to various lengths of the freely suspended monoatomic linear sodium wires (MLNaW), as nanojunctions between one-dimensional semi-infinite atomic leads of the same material. These calculations are motivated by the fact that the MLNaW nanojunctions, which have been investigated previously using other methods \cite{lee}, \cite{khomyakov}, \cite{egami2}, \cite{li}, \cite{egami3}, are known as good benchmark systems for electronic conductance calculations \cite{havu}, \cite{zugarramurdi}. This initial analysis is hence carried out in order to demonstrate the correctness and functionality of the PFMT method by comparing with previous results.

Further, our PFMT method is applied to the diatomic linear copper-cobalt wires (DLCuCoW) which are hitherto untreated. The particular interest in these systems stems from the fact that they are mechanically and thermodynamically stable at low temperatures, supported as they are on atomically flat Cu(111) surfaces \cite{lagoute1}, \cite{folsch}, \cite{lagoute2}. Our results for this nanojunctions supplement in a natural way recent experimental investigations on finite Cu-Co wires \cite{lagoute1}, by providing a theoretical description of the electronic transport properties for these systems.

The present paper is organized as follows. In section \ref{theory} we give an insight into the general and essential features of the PFMT approach. Section \ref{results} presents the applications of our model calculations to MLNaW and DLCuCoW nanojunctions. The numerical analysis yields both the transmission and the reflection probabilities for electrons, as well as the total electronic conductance for the considered systems at zero bias limit. The conclusions are presented in section \ref{summary}.

\section{THEORETICAL MODEL AND METHOD}
\label{theory}

The general formulation of the PFMT method is presented next. It is convenient to develop it initially, and to rewrite it afterwards into the form which is directly applied for our model calculations for the MLNaW and DLCuCoW nanojunctions. This allows us to introduce the PFMT technique in the most detailed and appropriate manner. Furthermore, the general presentation clarifies the mathematical approach which is applied in the present work for the system of wire nanojunctions. It equally illustrates how the method may be applied in general, treating general molecular forms for the nanojunction when conjugated between one-, two- or three-dimensional leads. The general formulation can also incorporate the interactions between atomic sites beyond nearest-neighbors.

\begin{figure}
\includegraphics[width=\columnwidth]{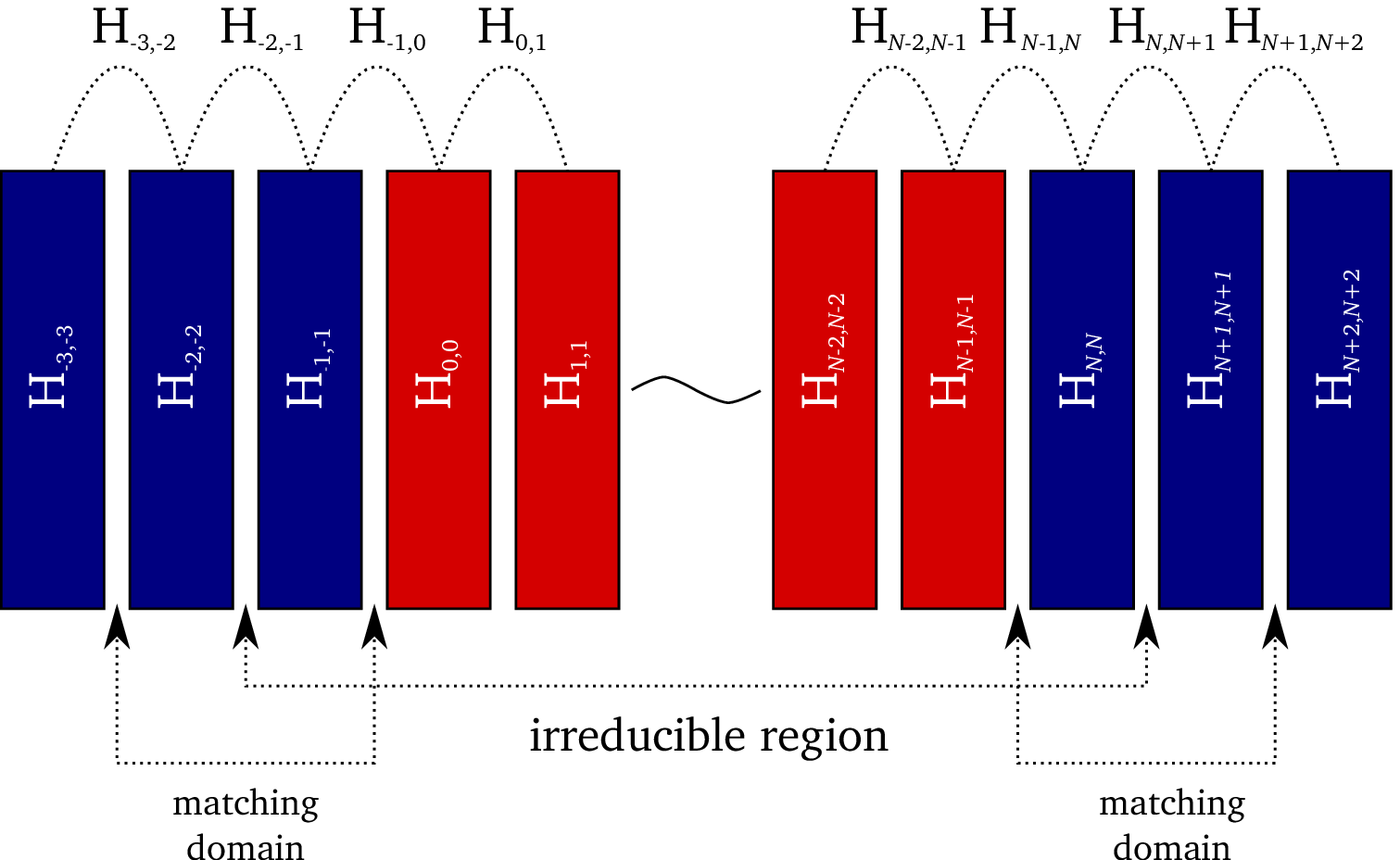}
\caption{Schematic planar projection in the direction of propagation of an arbitrary three-dimensional Landauer-B\"{u}ttiker-type system. The unit cells of the nanojunction and two semi-infinite leads are denoted by red and blue colors, respectively \cite{uc}. The corresponding Hamiltonian on- and off-diagonal sub-matrices are also depicted. Note that in order to keep the figure simple we show the Hamiltonian sub-matrices for interactions between different unit cells only for $n<n'$.}
\label{fig1}
\end{figure}

Let us first refer to the Landauer-B\"{u}ttiker approach for the analysis of the electronic transport via nanostructures \cite{landauer}, \cite{buttiker}, and divide an arbitrary nanojunction system into three main parts \cite{khater1}, \cite{uc}, namely the left and right leads, and the nanojunction domain itself, as is in Fig.\ref{fig1}.

To model the electronic properties of this system, the starting point is to write the general form of the tight-binding secular equation, for the total $M$ number of atoms per unit cell, and the total $L$ number of atomic orbitals per site, as follows
{\fontsize{7.8}{99}
\begin{eqnarray}
\label{eq1}
\nonumber
&E&\hspace{-7px}\left[ \begin{array}{c c c c c}
\ddots & \ddots &  &  &  \\
\ddots & \mathbf{S}_{n-1,n-1} & \mathbf{S}_{n,n-1}^{\dagger} &  &  \\
 &\mathbf{S}_{n,n-1} & \mathbf{S}_{n,n} & \mathbf{S}_{n+1,n}^{\dagger} &  \\
 &  & \mathbf{S}_{n+1,n} & \mathbf{S}_{n+1,n+1} & \ddots \\
  &  &  & \ddots & \ddots
\end{array} \right]\hspace{-6px}
\left[ \begin{array}{c}
\vdots  \\
\mathbf{c} (\mathbf{r}_{n-1}, \mathbf{k})  \\
\mathbf{c} (\mathbf{r}_{n}, \mathbf{k}) \\
\mathbf{c} (\mathbf{r}_{n+1}, \mathbf{k}) \\
\vdots
\end{array} \right]\\
&=&\hspace{-7px}\left[ \begin{array}{c c c c c}
\ddots & \ddots &  &  &  \\
\ddots & \mathbf{H}_{n-1,n-1} & \mathbf{H}_{n,n-1}^{\dagger} &  &  \\
 &\mathbf{H}_{n,n-1} & \mathbf{H}_{n,n} & \mathbf{H}_{n+1,n}^{\dagger} &  \\
 &  & \mathbf{H}_{n+1,n} & \mathbf{H}_{n+1,n+1} & \ddots \\
  &  &  & \ddots & \ddots
\end{array} \right]\hspace{-6px}
\left[ \begin{array}{c}
\vdots  \\
\mathbf{c} (\mathbf{r}_{n-1}, \mathbf{k})  \\
\mathbf{c} (\mathbf{r}_{n}, \mathbf{k}) \\
\mathbf{c} (\mathbf{r}_{n+1}, \mathbf{k}) \\
\vdots
\end{array} \right].
\end{eqnarray}
}
Note that we assume spin degeneracy in Eq.(\ref{eq1}). In what follows, the $M$ by $M$ overlap sub-matrix $\mathbf{S}_{n,n'}$ for a given $n$-th unit cell, incorporates the $s^{l,l'}_{m,m'}$ overlap integrals between the $l$ and $l'$ atomic orbitals centered at the sites corresponding to the real space vectors $\mathbf{r}_{m}$ and $\mathbf{r}_{m'}$, respectively. The unit cell may contain more than one inequivalent site, denoted here by $m$. Furthermore, the corresponding diagonal Hamitonian block matrix $\mathbf{H}_{n,n'}$ for the $n$-th unit cell, is composed of binding energy parameters $\varepsilon^{l}_{m}$ and the interaction integrals $h^{l,l'}_{m,m'}$. Then again, the off-diagonal $\mathbf{H}_{n,n'}$ sub-matrix contains only $h^{l,l'}_{m,m'}$ terms which describe interactions between different unit cells. Finally, the $\mathbf{c} (\mathbf{r}_{n}, {\mathbf{k}})$ denotes vector of the $c(\mathbf{r}_l-\mathbf{r}_m-\mathbf{r}_n, \mathbf{k})$ wave function coefficients which correspond to the $l$ atomic orbital at the lattice site $m$ in the $n$-th unit cell. $\mathbf{k}$ is the appropriate wave vector of the Bloch like wave function of the electronic excitation.

In order to calculate the electronic transport properties in the framework of the PFMT method for the nanoscale system presented in Fig.\ref{fig1}, we have to analyze the electronic scattering at the embedded nanostructure. To that end, we consider the nanojunction irreducible boundary domain, and the interactions of its electronic states with the electronic excitations incident from the leads to the left and right of the domain.
The wave function coefficients of consecutive unit cells in the leads are defined in the sense of the Bloch-Flouqet theorem by the following phase relation
\begin{equation}
\label{eq2}
\mathbf{c} (\mathbf{r}_{n}, \mathbf{k}) = z \mathbf{c} ( \mathbf{r}_{n-1}, \mathbf{k} ),
\end{equation}
where $z={\rm exp}(\pm i \mathbf{k} \mathbf{r}_{n})$ denotes the generalized Bloch phase factor. In our notation $z={\rm exp}(+ i \mathbf{k} \mathbf{r}_{n})$ describes the electronic wave function propagating to the right, and $z={\rm exp}(- i \mathbf{k} \mathbf{r}_{n})$ to the left. We should emphasize that in the case of multi-channel scattering processes one has to consider not only propagating but also evanescent waves \cite{khater1}. The evanescent solutions can be obtained using different procedures, however an elegant method presented previously for phonon and magnon problems \cite{khater1} shall be also applied to the electronic problem.

For an electron incident along the leads at a given energy $E$ and wave vector $\mathbf{k}$, where $E = E_{\gamma}(\mathbf{k})$ denotes the dispersion curves for the $\gamma$ = 1, 2,..., propagating waves in the Brillouin zone, the scattering at the boundary yields coherent reflected and transmitted fields. To calculate the Landauer-B\"{u}ttiker electronic conductance via a given nanojunction, we write the linearized equations of motion for the consecutive atomic sites within the irreducible nanojunction domain, on the basis of Eq.(\ref{eq1}). This procedure generates a set of $N+2M$ equations of motion with $N+4M$ unknown variables, namely the wave function coefficients, extending to outside the domain and inside the leads. Since the number of unknown variables is always greater than the number of equations, such a set of equations cannot be solved directly.

To resolve this problem, we use the phase field matching method \cite{khater1}, \cite{szczesniak}, \cite{khater2}. This implies that, for unit cells in the leads distant from the nanojunction boundary, the coefficients may be expressed in terms of an appropriate superposition of the eigenstates of the perfect leads at the same energy and wave vector. Hence, the wave function coefficients for the region outside the irreducible nanojunction domain may be expressed as
\begin{eqnarray}
\label{eq3}
\nonumber
\mathbf{c} (\mathbf{r}_{n}, \mathbf{k}) &=& \mathbf{c}_{\gamma} ( \mathbf{k} ) z^{-n}\\
&+& \sum^{\Gamma}_{\gamma'}  \mathbf{c}_{\gamma'} ( \mathbf{k} ) z^{n} r_{\gamma,\gamma'} \hspace{0.2cm} \textrm{for} \hspace{0.2cm} n \leqslant -1,
\end{eqnarray}
\begin{equation}
\label{eq4}
\mathbf{c} (\mathbf{r}_{n}, \mathbf{k}) = \sum^{\Gamma}_{\gamma'}  \mathbf{c}_{\gamma'} ( \mathbf{k} ) z^{n} t_{\gamma,\gamma'} \hspace{0.2cm} \textrm{for} \hspace{0.2cm} n \geqslant N.
\end{equation}
In equation (\ref{eq3}) and (\ref{eq4}), the $r_{\gamma,\gamma'}$ and $t_{\gamma,\gamma'}$ factors denote the reflection and transmission coefficients for the incident electronic wave function corresponding to the $\gamma$ channel, which is reflected or transmitted into the $\gamma'$ channel. Here, by channel we understand one of the total number $\Gamma$ of the allowed eigenstate solutions for the generalized Bloch phase factor in Eq.(\ref{eq2}), which determine the dispersion branches and the evanescent waves for the electronic structure of the lead system. The $\mathbf{c}_{\gamma} ( \mathbf{k} )$ term denotes hence the lead Hamiltonian eigenvector which corresponds to the $\gamma$ electronic eigenstate, and serves as a weighting factor in equations (\ref{eq3}) and (\ref{eq4}). For a complete description of the scattering processes it is essential to know the evanescent as well as the propagating electronic eigenstates \cite{khater1}.

Such formulation of the problem is similar to those presented in \cite{ando} and \cite{sorensen}. However, in contradiction to the mentioned techniques, we use equations (\ref{eq3}) and (\ref{eq4}) in order to directly rewrite the equations of motion for the region outside the nanojunction domain \cite{khater1}, \cite{szczesniak}, \cite{khater2}. In this manner we create an irreducible vector of the wave function coefficients for the nanojunction domain and its matching regions in the input and output leads, in the following way
\begin{equation}
\label{eq5}
\tilde{\mathbf{c}}(\mathbf{k})=[\mathbf{z}(\mathbf{k}) \times \mathbf{c}^{nano}(\mathbf{k})] +\mathbf{I}(\mathbf{k}),
\end{equation}
where $\mathbf{z}(\mathbf{k})$ is the $N+4M$ by $N+2M$ matrix composed of the phase factors which are mapped onto the atoms in the matching region outside the irreducible nanojunction domain. The vector $\mathbf{c}^{nano}(\mathbf{k})$ incorporates the vector of wave function coefficients on the nanojunction domain, and the reflection and transmission coefficients for the wave function in the matching region in the input and output leads. The number of reflection and transmission coefficients is equal to the number of scattering channels in an appropriately constructed Hilbert space. The vector $\mathbf{I}(\mathbf{k})$ decouples terms which describe the incoming wave. On the basis of Eq.(\ref{eq5}), we reduce next the resultant $N+2M$ set of equations for the irreducible region, to derive the following square matrix of inhomogenous linearized equations
\begin{equation}
\label{eq6}
[E\mathbf{S}-\mathbf{M}(z, \mathbf{k})] \times \mathbf{c}^{nano}(\mathbf{k})=-\tilde{\mathbf{I}}(\mathbf{k}).
\end{equation}
$\mathbf{S}$, the overlap matrix, and $\mathbf{M}$, composed of the Hamiltonian elements and appropriate phase factors, are $N+2M$ by $N+2M$ square matrices, whereas $\tilde{\mathbf{I}}(\mathbf{k})$ is the $\mathbf{I}(\mathbf{k})$ vector supplemented by the corresponding Hamiltonian terms. We observe that, contrary to the Green's function methods, Eq.(\ref{eq6}) allows us to solve the scattering problem directly for real energies.

Moreover, Eq.(\ref{eq6}) yields directly the $t_{\gamma,\gamma'}=t_{\gamma,\gamma'}(E)$ and $r_{\gamma,\gamma'}=r_{\gamma,\gamma'}(E)$ scattering amplitudes, and the wave function coefficients in the irreducible boundary domain, as a function of the electronic energy $E$ or wave vector $\mathbf{k}$. Here $E = E_{\gamma}(\mathbf{k})$ denotes the dispersion relation for the propagating waves of the electrons in the first Brillouin zone of the leads. It is these $E = E_{\gamma}(\mathbf{k})$ electrons, incident from the perfect leads, which scatter at the nanojunction.

In what follows, the $t_{\gamma,\gamma'}=t_{\gamma,\gamma'}(E)$ and $r_{\gamma,\gamma'}=r_{\gamma,\gamma'}(E)$  amplitudes yield the required Landauer-B\"{u}ttiker total transmission and reflection probabilities on the nanojunction, $T(E)$ and $R(E)$
\begin{equation}
\label{eq7}
R(E) = \sum_{\gamma,\gamma'} \frac{v_{\gamma'}}{v_\gamma} \left| r_{\gamma,\gamma'}(E) \right|^2,
\end{equation}
\begin{equation}
\label{eq8}
T(E) = \sum_{\gamma,\gamma'} \frac{v_{\gamma'}}{v_\gamma} \left| t_{\gamma,\gamma'}(E) \right|^2,
\end{equation}
summing over all incident, reflected, and transmitted waves. To ensure the unitarity character of the scattering matrix, the amplitudes are normalized with respect to their electronic group velocities, $v_{\gamma}$ and $v_{\gamma'}$, for electrons incident in channel $\gamma$ and reflected or transmitted in channel $\gamma'$. The unitarity condition, $T(E)+R(E)=1$, corresponds to the necessary conservation of energy.

Finally, assuming the zero bias limit and spin degeneracy, the overall conductance is then simply written as
\begin{equation}
\label{eq9}
G(N, E_F)=G_{0}T(E_F),
\end{equation}
where $G_{0}={{2e^2} / h}$ is the conductance quantum. The electronic conductance is given at the Fermi energy since the electrons at this level give the only significant contribution to the overall conductance from the Fermi-Dirac distribution.

\section{NUMERICAL RESULTS AND DISCUSSION}
\label{results}

\begin{figure}[t]
\centering
\includegraphics[width=\columnwidth]{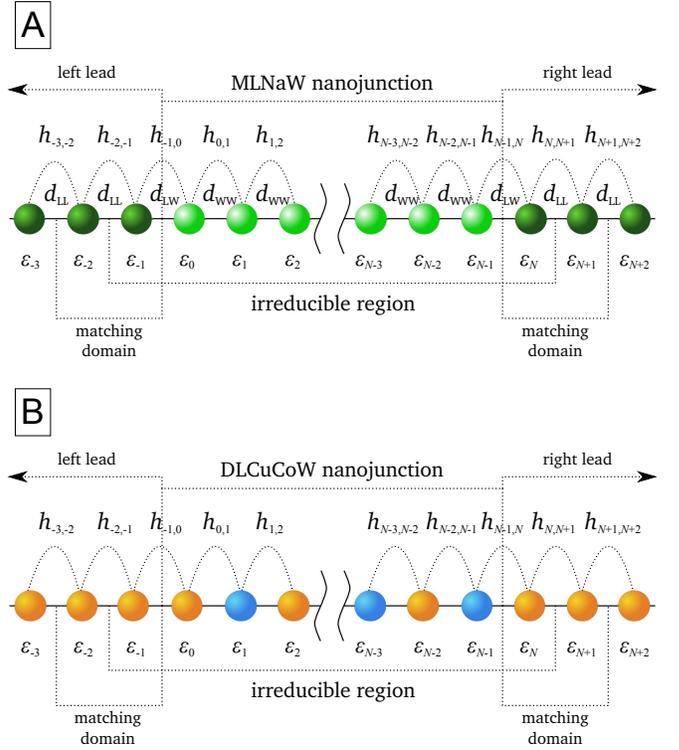}
\caption{Schematic representations of the: (A) $N$-atomic linear nanojunction made of sodium atoms (light green color) between sodium leads (dark green color); (B) $N$-atomic linear nanojunction made of pairs of copper (orange color) and cobalt (blue color) atoms between copper leads. The corresponding binding energies $\varepsilon_n$ for sites $n$, and the nearest-neighbor couplings $h_{n,n'}$ between sites $n$ and $n'$ are depicted. Note that in order to keep the figures simple we show the nearest-neighbor coupling terms only for $n<n'$. Additionally, in (A) three different interatomic spacings are considered: $d_{\rm WW}$ for the nanojunction wire, $d_{\rm LL}$ in the leads, and $d_{\rm WL}$ at the contact. The detailed description of the electronic and structural properties is presented in section \ref{results}.}
\label{fig2}
\end{figure}

In this section, we present the dynamic equations for the MLNaW and DLCuCoW nanojunctions on the basis of the theoretical discussion presented in section \ref{theory}. The schematic representations of the freely suspended MLNaW and the supported DLCuCoW nanojunctions between one-dimensional atomic electric leads are presented in Fig.\ref{fig2}(A) and Fig.\ref{fig2}(B), respectively. For both systems each of their atoms is characterized by a one valence $s$-type electronic state, and the interactions with adjacent atoms have a nearest-neighbor range, which is usually proposed for the Na atoms \cite{krans}. As for the considered Cu-Co systems, it has also been demonstrated, in view of previous experimental results \cite{lagoute1}, that their electronic dynamics can be described within the $s$-like monovalence approximation, suggesting that one channel is sufficient to model the electronic transmission.

For the numerical calculations of the electronic transport across MLNaW and DLCuCoW nanojuntions presented in this section, the overall Hamiltonian of the wire nanojunctions connected to the semi-infinite one-dimensional leads can be written on the basis of Eq.(\ref{eq1}) in the following reduced form
\begin{equation}
\label{eq10}
H = \sum_{n} \varepsilon_{n} + \sum_{n} h_{n,n'} (\delta_{n',n+1} + \delta_{n',n-1}),
\end{equation}
where we neglect the summation over different electronic states per site, and index $m$ is simply replaced by $n$ due to the fact that only one atom per unit cell is assumed in our calculations.

In the case of the MLNaW systems (see Fig.\ref{fig2}(A)), the Hamitonian elements are assumed to be consistent with the Harrison theory convention \cite{harrison}. In the presence of local charge neutrality, all the binding energies which characterize the atoms on the MLNaW nanojuctions are equal to each other, whereas the corresponding $h_{n,n'}$ parameters are distance dependent and defined as
\begin{equation}
\label{eq11}
h_{n,n'} \equiv \eta \frac{\hslash^2}{m_e d^{2}},
\end{equation}
where $\eta=-1.32$ is the dimensionless Harrison's coefficient for the $s$-type orbitals, $m_e$ denotes the electron vacuum mass, and $d$ depicts the nearest-neighbor interatomic distance. For the MLNaW nanojunctions, Eq.(\ref{eq11}) allows us to model the effective nearest-neighbor interactions by considering different interatomic spacings. On the other hand, the tight-binding parameters for DLCuCoW nanojunctions (see Fig.\ref{fig2}(B)) are assumed after \cite{lagoute1}, to be distance independent.

In particular, for the MLNaW, and DLCuCoW systems Eq.(\ref{eq6}) can be reduced to the following explicit form
\begin{eqnarray}
\label{eq12}
\nonumber
&&\left[ \begin{array}{c c c c c c}
h_{-1,-2} & -h_{-1,0} & 0 & \cdots & 0 \\
-h_{0,-1}z^{1} & E-\varepsilon_0 & \ddots &  &\vdots \\
0 & \ddots & \ddots & \ddots & 0 \\
\vdots &  & \ddots & E-\varepsilon_{N-1} & -h_{N-1,N}z^{N-1} \\
0 & \cdots & 0 & -h_{N,N-1} & h_{N,N+1}
\end{array} \right] \\
&&\times \left[ \begin{array}{c}
r \\
c ( 0, k )\\
\vdots \\
c [ (N-1)d, k ] \\
t\\
\end{array} \right]
= - \left[ \begin{array}{c}
h_{-1,-2} \\
-h_{0,-1}z^{-1}\\
0 \\
\vdots \\
0\\
\end{array} \right].
\end{eqnarray}
The total transmission and reflection probabilities, for the left and right leads, which are identical atomic wires presenting equal group velocities $dE/dk$ for the electrons, are related to the single channel amplitudes $t(E)$ and $r(E)$ in the following way
\begin{equation}
\label{eq13}
T(E)=|t(E)|^2 \hspace{0.7cm}{\rm and}\hspace{0.7cm} R(E)=|r(E)|^2.
\end{equation}
At this point we would like to draw attention to the fact that the PFMT formulation of the scattering problem presented in equations (\ref{eq3}-\ref{eq6}), allows us to introduce additional simplifications to the resultant equations. In particular, the boundary terms of matrix $[E\mathbf{I}-\mathbf{M}(z,k)]$ in Eq.(\ref{eq12}), known as a self-energies in the Green's function formalism, are real and energy independent in the PFMT method. This simplification can be also applied in the general case of Eq.(\ref{eq6}) when the interactions between neighbor atoms in the leads are symmetric.

Note, furthermore, that since the $s$-state electrons of the Na or Cu one-dimensional semi-infinite leads populate up to half of the available band states, one can show by symmetry that $E_F$ is equal to the binding energy of the lead atoms.

\subsection{Monatomic Sodium wire nanojunctions}

\begin{figure}
\includegraphics[width=\columnwidth]{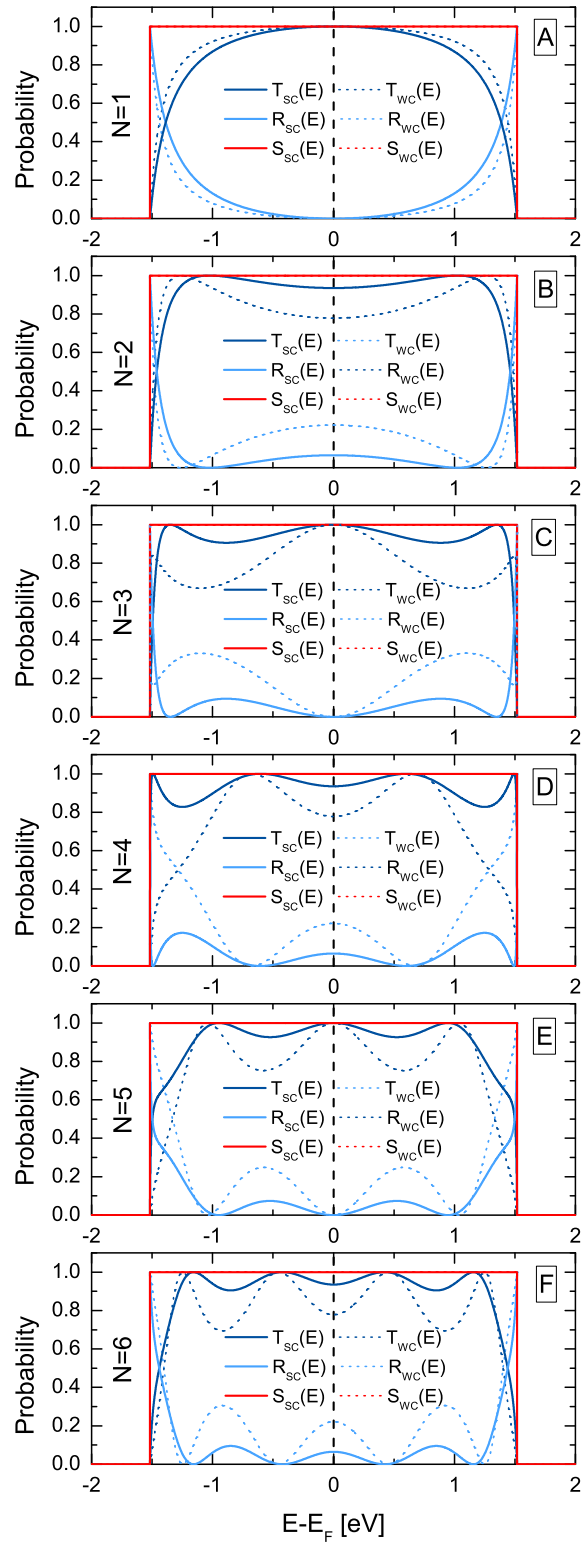}
\caption{The overall transmission $T(E)$ and reflection $R(E)$ probabilities as a function of energy for the MLNaW wire nanojunctions composed of $N$ atoms, with strong (SC) or weak (WC) lead-wire couplings. Subfigures (A), (C) and (E) correspond to the odd-number $N$, (B), (D) and (F) to the even-number. $E_F$ is set as a zero energy reference, and the unitarity condition is represented by the sum $S(E)$.}
\label{fig3}
\end{figure}

The PFMT model calculation is next applied numerically for the MLNaW wire nanojunctions between sodium leads as in Fig.\ref{fig2}(A). Our numerical results are determined for wires of various lengths, in particular they are presented typically here for $N\in [1,6]$.

As stated in section \ref{theory}, binding energies of Na atoms are constant and equal -4.96 eV, while the distance dependent coupling terms are modeled using Eq.(\ref{eq11}). The interatomic distance on the Na leads $d_{\rm LL} = 3.64 {\rm \AA}$ is considered to correspond to the bulk Na nearest-neighbor distance. However, $d$ of Eq.(\ref{eq11}) on the MLNaW nanojunction takes on a value of $d_{\rm WW} = 3.39 {\rm \AA}$ which is the average interatomic distance for the finite Na wires \cite{khomyakov}. In contrast, the lead-nanojunction spacing is considered for two numerical values, $d_{\rm WL}$ which is equal to $3.99 {\rm \AA}$ and 3.29 ${\rm \AA}$. These numbers correspond to a choice range of approximately $20\%$ weaker (WC) and stronger (SC) lead-nanojunction couplings, respectively, compared to the intrinsic lead coupling. This is a physically reasonable range to test. Taking into account the above values of the interatomic distances, we note that the $N$=1 case is equivalent to the situation of the infinite chain with the two bond defects, however this is not true for the remaining wire lengths due to the fact that $d_{\rm LL}\neq d_{\rm WW}$.

In Fig.\ref{fig3}, the overall transmission $T(E)$ and reflection $R(E)$ probabilities are presented as a function of the incident electronic energy $E$, for various wire lengths $N$, and for the two different lead-chain coupling values SC and WC defined above. For each sub-figure of Fig.\ref{fig3}, $T(E)$ and $R(E)$ are calculated independently, and the unitarity requirement for the scattering processes is checked by the sum $T(E)+R(E)$. Note that $E_F$ is set as a zero energy reference in these figures.

The positions of the transmission spectral resonances vary as a function of the lead-nanojunction couplings, as expected and as may be seen from Fig.\ref{fig3}. Due to the mirror symmetry and assumed local charge neutrality in the considered MLNaW systems, all energy levels, excluding the central peak for odd $N$, are distributed symmetrically with respect to the referential $E_F$, for both even and odd $N$. Analyzing the transmission spectra, the number of the resonances is always equal to the number of the atoms in the considered MLNaW wire up to $N=3$. This equality is also conserved for $N=4$ for the strong SC couplings, but reduced to $N-2$ for the weak WC couplings. The $N-2$ states are also observed for the five odd- and the six even-numbered wires for both SC and WC cases. The displacement of the resonances to outside the transmission spectra for any $N$ is directly related to the lead-nanojunction couplings for the considered system. This displacement happens when the resonance goes to a localized state outside the leads monovalent $s$-type conduction band. This effect is equally observed for resonances and localized vibration states on atomic nanojunctions \cite{khater1}. Further, when the MLNaW wire increases in length this displacement is quicker for the  WC connected wires than for the SC ones.

As mentioned previously, the only significant contribution to the overall conductance comes from electronic states in the neighborhood of the Fermi level $E_F$. It is hence noticeable that for the odd-number wires, one of the resonant maxima appears systematically at $E_F$, which yields as a consequence a conductance maximum for the corresponding systems. In contrast, the even-numbered wires do not present any maxima at $E_F$, which outcome results in successive minima for the conductance of these systems.

\begin{figure}[h!]
\centering
\includegraphics[width=\columnwidth]{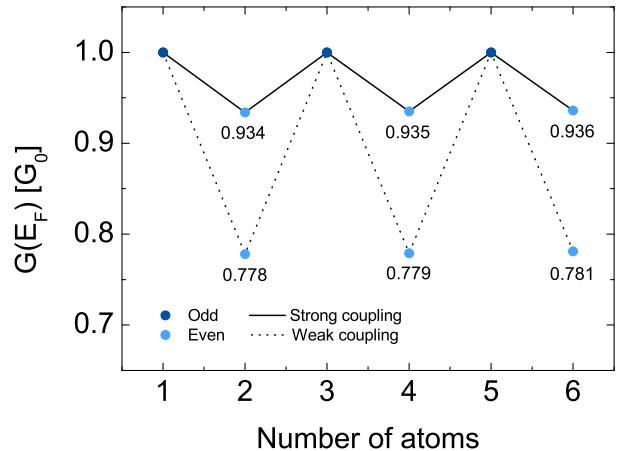}
\caption{The total electronic conductance $G(N, E_F)$ at the Fermi level as a function of the number of atoms on the MLNaW wire nanojunction, in units of $G_0$.}
\label{fig4}
\end{figure}

The total conductance $G(N, E_F)$ for the considered MLNaW wires is presented in Fig.\ref{fig4}, as a function of the wire length $N$, reflecting the maxima and minima of the transmission spectra at $E_F$, and exhibiting the usual conductance oscillations, here with a two-atom period. This situation relates directly to the interaction of the free electrons in the lead band with the electronic states of the nanojunction wire. In particular, since the $\sigma$ state for a sodium atom can take two electrons by the Pauli principle, one notes that there are half-empty states available for the odd-number wires  while non available for the even-number wires.  Since the Fermi energy $E_F$ is equal to the binding energy of the Na atoms, the odd-number wires present a greater local density of states LDOS accessible at the Fermi level $D(E_F)$, compared to that for the even-number wires. This ensures that the hopping mobility is unimpeded for the odd-number wires, whereas this is not the case for the even-number wires.  The corresponding electronic transport is consequently a maximum at $\sim G_0$ for the odd-number MLNaW wire nanojunctions, for both the weak WC and the strong SC couplings. The electronic conductance decreases for the even-number wires, and this decrease varies with the lead-nanojunction couplings, it is $\sim 0.065 G_0$ for SC and $\sim 0.221 G_0$ for WC, respectively. Furthermore, one observes that the conductance of the even-number MLNaW wire nanojunctions increases slightly but monotonically with the increase of the wire length. This situation may be explained qualitatively by arguing that the electric leads have a bigger influence on the LDOS at the Fermi level for the short wires than for the longer ones. An interesting discussion on the problem of how the different types of leads may influence the overall electronic conductance can be found in \cite{khomyakov}.

In summary, the electronic conductance determined in this subsection presents hitherto unknown detailed transmission spectra, and confirms the well known even-odd conductance oscillation effect \cite{yamaguchi}, which is previously reported for the MLNaW nanojunctions \cite{lang1}, as well as for other wire systems \cite{lee}, \cite{thygesen}, \cite{lang2}. We also note that our results are in agreement with those obtained on the basis of the first-principle calculations in other model calculations \cite{lee}, \cite{egami2}, \cite{egami3}. This clearly confirms the validity of the PFMT approach. The minor quantitative differences between our results for the conductance and those previously reported, for the same $N$-number MLNaW wire systems, appear mainly due to the assumed structures of the electric leads.

\subsection{Diatomic Copper-Cobalt wire nanojunctions}

\begin{figure*}[ht]
\centering
\includegraphics[width=\textwidth]{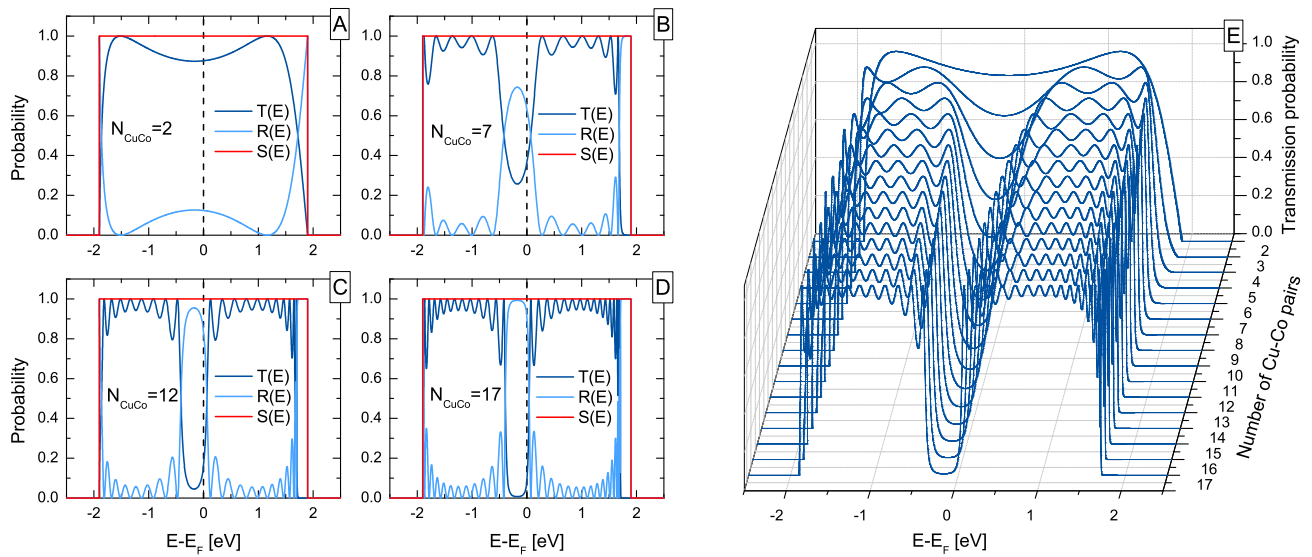}
\caption{(A) - (D): The selected spectra of the transmission $T(E)$ and reflection $R(E)$ probabilities for respectively 2, 7, 12 and 17 copper-cobalt atomic pairs in the DLCuCoW scattering region. The unitarity condition is represented by the sum $S(E)$. (E): Surface plot of the transmission probabilities as a function of energy and number of the Cu-Co atomic pairs in the DLCuCoW scattering region. For all subfigures the $E_F$ is set at zero of energy.}
\label{fig5}
\end{figure*}

The second considered system in this work concerns the supported DLCuCoW wire nanojunctions presented in section II. Finite - Cu - Co - wire systems have been experimentally prepared previously as grouped adatoms on the flat Cu(111) surface by Lagoute {\it et al.} \cite{lagoute1}, with the use of the low-temperature scanning tunneling microscope technique. Initially, these systems have been developed as magnetic/nonmagnetic finite atomic wires for the investigation of their potential magnetic properties. Subjects concerning their spin dynamics \cite{saubanere}, their storage potential \cite{brovko}, and the observation of the Kondo effect in their structure \cite{neel}, have been indeed addressed recently. However, their electronic transport properties have not been considered previously, despite their intrinsic interest under a  DLCuCoW configuration, and their mechanical and thermodynamical stability at low  temperatures. Note in this respect the increasing interest in the stability of such supported wire systems \cite{oncel} for arbitrary ranges of temperature.

As it was stated in section II, the electron dynamics of the supported DLCuCoW nanojunctions can be discussed within the $s$-like single band effective tight-binding model proposed in \cite{lagoute1}. This possibility arises from the existence of unoccupied confined quantum states formed along the Cu-Co wire nanojunctions by linking the orbitals of $sp_z$ character. Such confined states have been observed in other supported wire systems like the monatomic Cu \cite{folsch} and Au \cite{nilius}, the diatomic Au-Pd nanojunction systems \cite{wallis}, and Fe nano-islands \cite{delga}.

In this context the possibility of using semiconductor surfaces which present a band gap is equally interesting as a support for the atomic wires. These surfaces may electronically decouple from the supported wire orbitals in certain cases, allowing one to study the electronics of stable one-dimensional wire systems \cite{oncel}.

From our point of view the diatomic nanojunction wires may be used to control the value of the overall electronic conductance in low-dimensional nanoelectronic circuits. In particular, the foreign cobalt atoms injected directly into the monatomic copper wire, act as controllable chemical defects which modify the initial system and its electronic properties as a consequence. Furthermore, the substrate supported configurations of the lead - DLCuCoW wire nanojunctions - lead, suggest a greater mechanical and thermodynamic stability than for the freely suspended wires, which is one of the most important features required for future nanoelectronic devices.

Our model calculations for the electronic quantum conductance of the DLCuCoW systems are presented for wires composed of different numbers of periodically arranged Cu-Co atomic pairs, as in Fig.\ref{fig2}(B). The length of the DLCuCoW wire may characterized by the number $N_{\rm CuCo}$ of such pairs, and the results in this subsection are hence presented in terms of this latter variable for a given DLCuCoW nanojunction.

In the case of the DLCuCoW nanojunctions, we assume the Hamiltonian elements of Eq.(\ref{eq10}) to have the values proposed in \cite{lagoute1}. Following these studies, the binding energy 3.31 eV is considered to be the same for all the copper atoms, whereas for cobalt it is 2.96 eV. The coupling terms between copper nearest-neighbors on the leads, and between copper and cobalt nearest-neighbors on the wire nanojunction, are close to each other with values of $-0.95$ eV and $-0.94$  eV, respectively.

It is important to note that the $N_{\rm CuCo}= 1$ nanojunction wire is a very particular case since it has quite a different symmetry from all other lengths and is open to different structural definitions; we have calculated the total conductance for this particular case to be $0.9658G_0$, but will not use it in our discussion of the properties of the conductance $G(N_{\rm CuCo}, E_F)$ for a general wire length.

In Fig.\ref{fig5}(A)-(D) we present the detailed transmission $T(E)$ and reflection $R(E)$ spectra of the DLCuCoW nanojunction, for the selected values of $N_{\rm CuCo}$ = 2, 7, 12, and 17. The resonance maxima and minima can be observed as for the MLNaW wires, where $E_F$ is again set as a zero energy reference. However, the transmission and reflection spectra are not symmetric with respect to this reference, due the slight difference between the cobalt and copper binding energies. The most important  conclusion concerns the value of the transmission probabilities close to $E_F$. The transmission spectra in Fig.\ref{fig5}(A)-(D) show strong resonant minima for both the even- and the odd-number wires. In particular, we observe that the value of the transmission at the Fermi level decreases when the length of the effective nanojunction increases. This trend is summarized in Fig.\ref{fig5}(E) for the ensemble of the transmission spectra for $N_{\rm CuCo}$ values ranging from 2 to 17. At the upper limit the conductance goes to $G(N_{\rm CuCo}=17, E_F)\sim 10^{-1}G_0$ which is sufficient for the purpose of the present calculation.

\begin{figure}[h]
\centering
\includegraphics[width=\columnwidth]{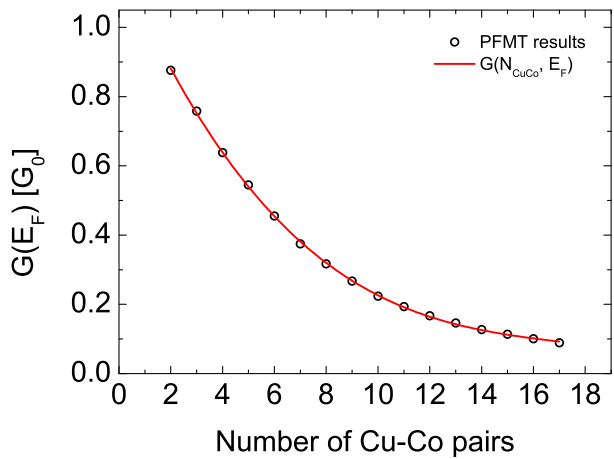}
\caption{The total electronic conductance $G(N_{\rm CuCo}, E_F)$ in units of $G_0$, read at the Fermi level $E_F$ as a function of the number of Cu-Co pairs on the DLCuCoW nanojunction wire. The PFMT results are represented by open circles, and the fitting function by the red curve.}
\label{fig6}
\end{figure}

The overall electronic conductance $G(N_{\rm CuCo}, E_F)$ as a function of the DLCuCoW wire length in terms of Cu-Co atomic pairs, is presented in Fig.\ref{fig6}, along with a fitting exponential function
\begin{eqnarray}
\label{eq14}
\nonumber
&&G(N_{\rm CuCo}, E_F)=G(2, E_F){\rm exp}[-\alpha (N_{\rm CuCo}-2)]\\
&&{\rm for} \hspace{0.2cm} N_{\rm CuCo} \geq 2.
\end{eqnarray}
Fig.\ref{fig6} resumes the transmission spectra at the Fermi energy, it shows the strong monotonic decay of the electronic conductance with the increase of the number of Cu-Co pairs of the nanojunction. In the corresponding Eq.(\ref{eq14}), $G(2, E_F)=0.8760G_0$ is the conductance of a wire made up of two Cu-Co pairs, and $\alpha=0.1630$ denotes the decay constant. Eq.(\ref{eq14}) allows one hence to read the numerical values of the conductance $G(N_{\rm CuCo}, E_F)$ of the DLCuCoW wires for $N_{\rm CuCo} \geq 2$ using a simple formula which is useful in the analysis of potential applications of corresponding nanocircuits. This behavior is fundamentally different from the typical conductance oscillations observed for odd- and even-number monoatomic wires.

\begin{figure}[h]
\centering
\includegraphics[width=\columnwidth]{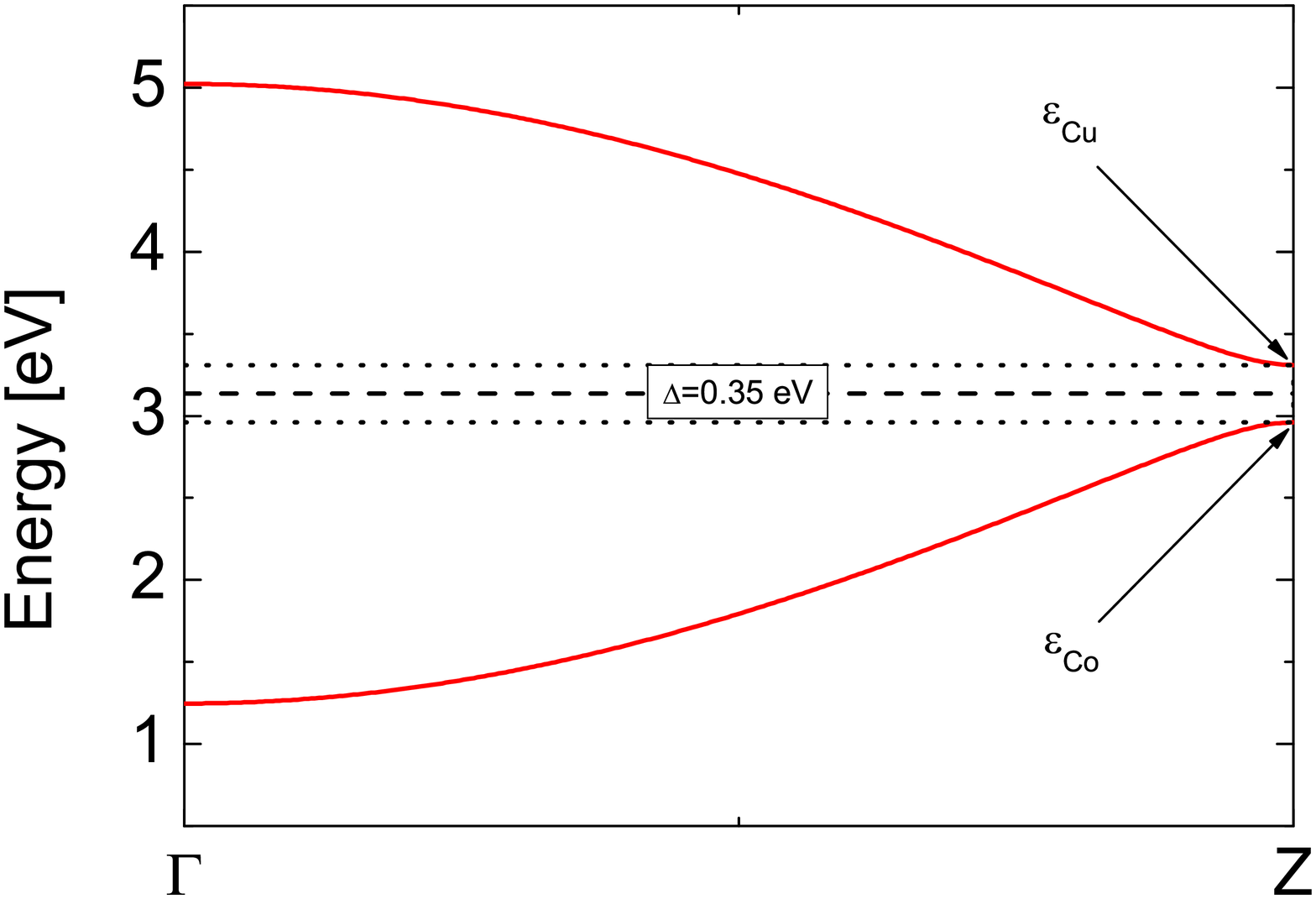}
\caption{Band structure of the infinite variant of the DLCuCoW wire nanojunction over the first Brillouin zone. The band gap (${\it \Delta}$=0.35 eV) is marked by two dotted lines, between the copper ($\varepsilon_{Cu}$) and the cobalt ($\varepsilon_{Co}$) binding energies. The Fermi energy for the half-filled bands is represented by the dashed line.}
\label{fig7}
\end{figure}

To understand the decrease of the conductance in Fig.\ref{fig6}, one has to take a closer look at the band structure of the infinite -Cu-Co- diatomic chain which is the infinite limit of the DLCuCoW nanojunction. We have calculated the band structure for such an infinite diatomic wire as in Fig.\ref{fig7}. This presents a band gap of $\it {\Delta}$=0.35 eV and renders the system  theoretically insulating. This band gap, which corresponds directly to the difference of the binding energies for cobalt and copper atoms, presents a differential $\sim 10^{-1}$ with respect to the Cu binding energy. It is an effective potential barrier for electrons between successive Cu and Co sites, and results in a small but cumulative decrease of the conductance as the DLCuCoW wire nanojunction increases in length, when comparing to the theoretical maximum of $G_0$ for the pure atomic Cu wire. This direct interpretation is confirmed by the exponential form of Eq.(\ref{eq14}), where exp$[-\alpha (N-2)]$ = exp$(-\alpha)^{N-2}$, expressing the monotonic decay of the electronic conductance with the increasing length of the DLCuCoW nanojunction.

\section{SUMMARY}
\label{summary}

In this work we present a general method based on the algebraic phase field matching theory (PFMT), to calculate the electronic quantum transport across nanojunctions of arbitrary form between material leads of $n$-dimensions, $n$ = 1, 2 or 3. This PFMT method is applied in particular to calculate the quantum electronic conductance across wire nanojunctions held between one-dimensional electric leads. The presented PFMT formalism provides a compact and efficient approach for the analysis of the electronic quantum transport for a wide range of wire nanojunctions \cite{szczesniak2}.

The model calculations are carried out for nanojunctions made up of sodium atomic wires (MLNaW) between leads of the same element. Our calculated results for the electronic quantum transport are shown to be a function of the physical parameters of the wires, and exhibit the well known oscillation behavior for even- and odd-number wire lenghts. They are also in agreement with the first-principle results \cite{lee}, \cite{egami2}, \cite{egami3}. This clearly confirms the validity of the presented PFMT model.

Our model calculations are also carried out for the diatomic copper-cobalt wires (DLCuCoW) held between copper leads. This system is selected for its mechanical and thermodynamical stability at low temperatures, which is one of the important features required for future nanoelectronic devices. In contrast to the MLNaW system, the electronic quantum transport of the  DLCuCoW wire nanojunction, at the Fermi energy, exhibits exponential decay with increasing wire length. This behavior is explained notably in terms of the band gap of the infinite variant of the DLCuCoW nanojunction wire. A relation, depicting analytically the numerical values of the conductance $G(N_{\rm CuCo}, E_F)$ of the DLCuCoW wires for any length $N_{\rm CuCo} \geq 2$, is given and may help to simplify the analysis of potential applications for corresponding nanocircuits.

The presented PFMT approach, a transparent and time-saving method, should be especially interesting for the treatment of complex systems presenting multichannel conductance, and for the treatment of the electronic quantum transport of nanojunctions which can exhibit electron-electron and electron-phonon interactions at high temperatures or due to structural disorder.

\section{Acknowledgments}

D. Szcz\c{e}\'{s}niak would like to note that this work has been financed by the Polish National Science Center (grant DEC- 2011/01/N/ST3/04492). He also acknowledges support from the French Ministry of Foreign Affairs, and the University of Maine 3MPL Graduate School. The authors would like to thank Pr Z. B\c{a}k for his kindness and support throughout this work, and Pr W.A. Harrison, Dr R. Szcz\c{e}\'{s}niak, and Dr J. Lagoute for useful discussions.


\end{document}